\documentstyle[preprint,aps,prabib]{revtex}

\begin{document}

\title{Quantization of classical 
singular solutions in Yang-Mills theory}

\author{V. Dzhunushaliev
\thanks{E-mail address: dzhun@rz.uni-potsdam.de, 
dzhun@freenet.bishkek.su}}
\address{Universit{\"a}t Potsdam, Institute f{\"u}r Mathematik,
14469, Potsdam, Germany \\
and Theor. Physics Dept., KSNU, 
720024, Bishkek, Kyrgyzstan}
\author{D. Singleton
\thanks{E-mail address : dougs@csufresno.edu}}
\address{Dept. of Physics, CSU Fresno, 2345 East San Ramon Ave.,
M/S 37, Fresno, CA 93740-8031}
\date{\today}

\maketitle

\begin{abstract}
In this paper we apply a variant of
Heisenberg's quantization method for
strongly interacting, non-linear fields,
to solutions of the classical Yang-Mills
field equations which have bad asymptotic behavior.
After quantization we find that the bad features ({\it i.e.} divergent
fields and energy densities) of these solutions
are moderated. From these results we argue that in
general the n-point Green's functions for Yang-Mills
theories can have non-perturbative pieces which
can not be represented as the sum of Feynman diagrams. 
A formalism for dealing with these non-Feynman pieces
via nonassociative field operators is suggested. These
methods may also find some application in dealing
with high-$T_c$ superconductors.
\end{abstract}
\narrowtext

\section{Introduction}

The quantization of strongly coupled, nonlinear fields 
is a difficult and still open problem. The
success of QED and the Standard Model of electroweak
interactions is largely due to the small size 
of the coupling constants, which allows for the
perturbative expansion of physical quantities in terms of a power
series in the coupling constants. This Feynman diagrammatic
technique is the standard method for
dealing with weakly coupled theories. For strongly
interacting field theories such as quantum chromodynamics
(QCD) this procedure can not be applied in energy
regimes where the coupling constant is large. The theoretical
demonstration of confinement and the
determination of the mass spectrum of light mesons and
baryons are two examples where the perturbative
techniques of quantum field theory are not good tools.
These low energy QCD phenomenon must be dealt with
as completely nonperturbative, quantum effects. 
\par 
A mathematical way of stating this basic difficulty in QCD
is that one needs some nonperturbative method
for calculating the n-point Green functions of the
theory. Usually these problems in strongly coupled
field theories are dealt with using numerical simulations,
but the computing power of modern
computers does not yet permit, in general, the
same kind of precise comparison between theory and
experiment as is found in QED or the electroweak theory.
Even if the state of the art in lattice gauge
theory calculations increases to the same level of
precision, it would still be desirable
to have approximate, analytical methods for investigating
these non-perturbative aspects of QCD or any strongly coupled field theory.
Here we apply a nonperturbative method for calculating
the n-point Green's functions of a theory which was
originally proposed in the $50 's$ by Heisenberg 
(see Refs. \cite{hs1}, \cite{hs2} and \cite{hs3})
to deal with the strongly coupled, non-linear Dirac
equation. In this method one writes down an infinite
system of equations which connects together all the
n-point Green's functions of the theory. Then by
introducing some physically reasonable approximations
one reduces this infinite system of equations
into a finite system. Here we apply this quantization
procedure to several classical solutions in 
SU(2) and SU(3) gauge theories. Although these
classical solutions possess some interesting properties
they also have bad features which are eliminated
or reduced after the Heisenberg quantization is applied.

\section{Classical singular solutions in SU(2) 
and SU(3) gauge theories}

In this section we examine several classical solutions 
for SU(2) and SU(3) Yang-Mills gauge theories. These solutions 
all have similar, qualitative asymptotic behavior: 
one part of the classical gauge potential diverges as a
power law and another  part oscillates strongly.
In this section we follow the notation and
conventions of Ref. \cite{dzh1}.

\subsection{The SU(2) gauge ``string''}

Consider the following cylindrical ansatz for
SU(2) Yang-Mills theory 
\begin{mathletters}
\label{16}
\begin{eqnarray}
A^1_t & = & f(\rho ),
\label{16:1}\\
A^2_z & = & v(\rho ),
\label{16:2}\\
A^3_{\phi} & = & \rho w(\rho ),
\label{16:3}
\end{eqnarray}
\end{mathletters}
$z, \rho , \phi$ are the standard cylindrical coordinates;
$a=1,2,3$ are SU(2) color indices. Inserting
this ansatz into the Yang - Mills field equations
\begin{equation}
\frac{1}{\sqrt{-g}}\partial _{\mu} \left (\sqrt {-g} {F^{a\mu}}_{\nu}
\right ) + f^{abc} {F^{b\mu}}_{\nu} A^c_{\mu} = 0,
\label{2}
\end{equation}
yields
\begin{mathletters}
\label{17}
\begin{eqnarray}
f'' + \frac{f'}{\rho} & = & f\left (v^2 + w^2 \right ),
\label{17:1}\\
v'' + \frac{v'}{\rho} & = & v\left (-f^2 + w^2 \right ),
\label{17:2}\\
w'' + \frac{w'}{\rho}  - \frac{w}{\rho ^2}& = & w
\left (-f^2 + v^2 \right ),
\label{17:3}
\end{eqnarray}
\end{mathletters}
In the case where $w=0$ Eqs. (\ref{17}) reduces to 
\begin{mathletters}
\label{18}
\begin{eqnarray}
f'' + \frac{f'}{\rho} & = & fv^2 ,
\label{18:1}\\
v'' + \frac{v'}{\rho} & = & -vf^2 .
\label{18:2}
\end{eqnarray}
\end{mathletters}
The asymptotic behavior of the ansatz functions $f, v$ and the energy 
density ${\cal E}$ are
\begin{mathletters}
\label{21}
\begin{eqnarray}
f & \approx & 2\left [x + \frac{\cos \left (2x^2 + 2\phi _1\right )}
{16x^3} \right ] ,
\label{21:1}\\
v & \approx & \sqrt{2} \frac{\sin \left (x^2 + \phi _1 \right )}{x} ,
\label{21:2}\\
{\cal E} & \propto & f'^2 + v'^2 + f^2v^2 \approx const,
\label{21:3}
\end{eqnarray}
\end{mathletters}
where $x=\rho /\rho _0$ is a dimensionless 
radius, and $\rho _0, \phi _1$ 
are  constants. A numerical study of 
Eqs. (\ref{18}) was carried out in Ref. \cite{dzh1},
and it was found that $A^1_t=f(\rho)$
was a confining, power law potential and 
$A^2_z=v(\rho)$ was a strongly oscillating 
potential . Depending on the relationship between 
the initial conditions $v(0)$ and $f(0)$ 
the energy density near $\rho = 0$ was either a hollow 
({\it i.e.} an energy density less than the asymptotic value)
or a hump ({\it i.e.} an energy
density greater than the asymptotic value).
On account of this and the cylindrical symmetry of this solution 
we called this the ``string'' solution.

\subsection{SU(3) ``bunker'' solution}
\label{su3}

Next we considered the following SU(3) ansatz for the
gauge fields \cite{palla} \cite{gal} \cite{pagel} :
\begin{mathletters}
\label{1}
\begin{eqnarray}
A _0 & = & \frac{2\varphi(r)}{{\bf i} r^2} \left( \lambda ^2 x - \lambda ^5 y 
      + \lambda ^7 z\right ) + \frac{1}{2}\lambda ^a
      \left( \lambda ^a _{ij} + \lambda ^a_{ji} \right ) 
      \frac{x^ix^j}{r^2} w(r),
\label{1a:1}\\
A^a_i & = & \left( \lambda ^a_{ij} - \lambda ^a_{ji} \right )
        \frac {x^j}{{\bf i} r^2} \left(f(r) - 1\right ) +
        \lambda ^a_{jk} \left (\epsilon _{ilj} x^k + 
        \epsilon _{ilk} x^j\right ) \frac{x^l}{r^3} v(r)
\label{1a:2},
\end{eqnarray}
\end{mathletters}
here $\lambda ^a$ are the Gell - Mann matrices; 
$a=1,2,\ldots ,8$
is a color index; the Latin indices $i,j,k,l=1,2,3$ 
are space indices; 
${\bf i}^2=-1$; $r, \theta, \phi$ are 
spherically coordinates. 
Substituting Eqs. (\ref{1}) into the Yang - Mills 
equations (\ref{2}) 
gives the following system of equations for $f(r), v(r), w(r)$ 
and $\varphi (r)$ 
\begin{mathletters}
\label{3}
\begin{eqnarray}
r^2f''& =& f^3 - f + 7fv^2 + 2vw\varphi - f\left (w^2 + \varphi ^2\right ),
\label{3a:1}\\
r^2v''& = & v^3 - v + 7vf^2 + 2fw\varphi - v\left (w^2 + \varphi ^2\right ),
\label{3b:2}\\
r^2w''& = & 6w\left (f^2 + v^2\right ) - 12fv\varphi,
\label{3c:3}\\
r^2\varphi''& = & 2\varphi\left (f^2 + v^2\right ) - 4fvw.
\label{3d:4}
\end{eqnarray}
\end{mathletters}
This set of equations is difficult to solve even numerically
so we investigated simplified cases. 
\par 
First we examined the $f=\varphi=0$ case, under which 
Eqs. (\ref{3}) became
\begin{mathletters}
\label{9}
\begin{eqnarray}
r^2 v''& = & v^3 - v - vw^2,
\label{9:1}\\
r^2 w''& = & 6w v^2.
\label{9:2}
\end{eqnarray}
\end{mathletters}
In the asymptotic limit $r \rightarrow \infty$ 
the form of the solution 
to Eqs. (\ref{9}) approaches a form similar to 
the ``string'' solution: 
\begin{mathletters}
\label{11}
\begin{eqnarray}
v & \approx & A \sin \left (x^{\alpha } + \phi _0\right ),
\label{11:1}\\
w & \approx & \pm\left [ \alpha  x^{ \alpha } + 
\frac{\alpha -1 }{4}\frac{\cos {\left (2x^{\alpha} + 2\phi _0 \right )}}
{x^{\alpha}}\right ],
\label{11:2}\\
3A^2 & = & \alpha(\alpha - 1).
\label{11:3}
\end{eqnarray}
\end{mathletters}
where $x=r/r_0$ is a dimensionless radius and $r_0, \phi _0$, and $A$ are 
constants. A numerical investigation of Eqs. (\ref{9}) 
was done in Ref. \cite{dzh1}. Since $v(r)$ was
strongly oscillating this resulted in the
space part of the gauge field of Eq. (\ref{1a:2}) being strongly oscillating.
The ansatz function $w(r)$ increased as some power of $x$ as $x \rightarrow
\infty$, which indicated a classically confining potential.
\par  
The ``magnetic'' and ``electric'' fields associated with this solution
can be found from $A_{\mu} ^a$, and have the following behavior 
\begin{mathletters}
\label{12}
\begin{eqnarray}
H^a _r & \propto & \frac{v^2-1}{r^2} , \; \; \; \; \; \; \; \;
H^a_{\phi}  \propto  v' , \; \; \; \; \; \; \; \;
H^a_{\theta}  \propto  v' , \\
\label{12:1}
E^a_r & \propto & \frac{rw' - w}{r^2}, \; \; \; \; \; \; \; \;
E^a_{\phi} \propto  \frac{vw}{r}, \; \; \; \; \; \; \; \; 
E^a_{\theta}  \propto  \frac{vw}{r},
\label{12:2}
\end{eqnarray}
\end{mathletters}
for $E^a _r , H^a _{\theta}$, and $H^a_{\phi}$ the color index 
is $a=1,3,4,6,8$ and for $H^a _r, E^a _{\theta}$ and $E^a _{\phi}$ 
$a=2,5,7$. The asymptotic behavior of 
$H^a_{\phi}, H^a_{\theta}$ and $E^a_{\phi}, E^a_{\theta}$ 
is dominated by the strongly oscillating function $v(r)$. 
Later we will show that the quantum corrections to this solution 
tend to smooth out  these strongly oscillating fields. 
>From Eqs. (\ref{12}) and the asymptotic form (\ref{11}) 
the radial components of the ``magnetic'' and ``electric'' 
have the following asymptotic behavior
\begin{equation}
\label{13}
H^a_r  \propto  \frac{1}{r^2}, \; \; \; \; \; \;
E^a_r  \propto  \frac{1}{r^{2-\alpha}}.
\end{equation}
The radial ``electric'' field falls off slower than 
$1/r^2$ (since $\alpha > 1$) indicating the presence of a confining
potential. The $1/r^2$ fall off of $H^a _r$ indicates that this solution
carries a ``magnetic'' charge. This was also true for the solutions
discussed in Refs. \cite{palla} \cite{pagel}. 
\par 
This solution of the classical SU(3) field equations
exhibits fields which lead to a classical confining
behavior, and which have similarities with certain phenomenological
potential models of confinement. The most significant
draw back of these solutions is that they have infinite
field energy. In the case of the ``bunker'' solution one
finds that the asymptotic form of the energy density goes as
\begin{equation}
{\cal E} \propto 4\frac{v'^2}{r^2} + \frac{2}{3}\left (
\frac{w'}{r} - \frac{w}{r^2}\right ) ^2 + 4\frac{v^2w^2}{r^4} +
\frac{2}{r^4} \left (v^2 - 1\right )^2 \approx \frac{2}{3}
\frac{\alpha ^2 (\alpha -1) (3 \alpha -1)}{x^{4-2\alpha}}
\label{14}
\end{equation}
As $\alpha > 1$ this energy density will yield an
infinite field energy when integrated over all space. This can
be compared with the finite field energy monopole and dyon solution
\cite{bps}. However it has been
demonstrated \cite{swank} that the finite energy monopole solutions
can not trap a test particle while the infinite energy solutions
can.

\subsection{The classical f, $\varphi \neq 0$ case}

Finally we examine the $v=w=0$ case:
\begin{mathletters}
\label{monp1}
\begin{eqnarray}
r^2f'' & = & f^3 - f - f\varphi ^2, 
\label{monp1:1} \\
r^2\varphi '' & = & 2\varphi f^2
\label{monp1:2}
\end{eqnarray}
\end{mathletters}
These equations have well known, nonsingular 
monopole solutions, but these monopole solutions require 
special boundary conditions. If one does not impose
special conditions then a more general solution is
\begin{mathletters}
\label{monp4}
\begin{eqnarray}
f & \approx & A \sin \left (x^{\alpha } + \phi _0\right ),
\label{monp4:1}\\
\varphi & \approx & \pm\left [ \alpha  x^{ \alpha } + 
\frac{\alpha -1 }{4}\frac{\cos {\left (2x^{\alpha} + 2\phi _0 \right )}}
{x^{\alpha}}\right ],
\label{monp4:2}\\
A^2 & = & \alpha(\alpha - 1).
\end{eqnarray}
\end{mathletters}
which is the same as the asymptotic form of the
``bunker'' solution, and thus shares its bad long
distance behavior. A numerical investigation of the
solutions to Eq. (\ref{monp1}) also indicates a similarity to
the ``bunker'' solution.
The ``magnetic'' and ``electric'' fields
associated with this solution are also similar to those
of the ``bunker'' solution.
\begin{mathletters}
\label{monp3}
\begin{eqnarray}
H^a _r & \propto & \frac{f^2-1}{r^2} , \; \; \; \; \; \; \; \;
H^a_{\phi}  \propto  f' , \; \; \; \; \; \; \; \;
H^a_{\theta}  \propto  f' , \\
\label{monp3:1}
E^a_r & \propto & \frac{r\varphi ' - \varphi }{r^2}, 
\; \; \; \; \; \; \; \;
E^a_{\phi} \propto  \frac{f\varphi }{r}, \; \; \; \; \; \; \; \; 
E^a_{\theta}  \propto  \frac{f\varphi }{r},
\label{monp3:2}
\end{eqnarray}
\end{mathletters}
here the color indices $a=2,5,7$ indicate that this is
really an $SU(2)$ gauge potential embedded within SU(3).
We now review the Heisenberg quantization method in order to apply it
to these three solutions.

\section{Heisenberg Quantization of Strongly Interacting Fields}

Heisenberg's basic idea is that the n-point Green's
functions can be found from some infinite set of coupled
differential equations, which are
derived from the field equations for the field operators. As an example
we show how this method of quantization works for a spinor field with
a nonlinear self interaction \cite{hs1} \cite{hs2} \cite{hs3}. 
The basic equation for the spinor field is :
\begin{equation}
\label{h1}
\gamma ^{\mu} \partial _{\mu} {\hat \psi} (x) - 
l^2 \Im [{\hat \psi}(x) ({\hat {\bar \psi}}(x) 
{\hat \psi}(x)) ] = 0
\end{equation}
where $\gamma ^{\mu}$ are Dirac matrices; 
${\hat \psi} (x), {\hat {\bar \psi}}(x)$ are 
the operators of the spinor field and its adjoint respectively; 
$\Im [{\hat \psi} ({\hat {\bar \psi}} 
{\hat \psi)} ] = {\hat \psi} ({\hat {\bar \psi}} 
{\hat \psi})$ or ${\hat \psi} \gamma ^5 
({\hat {\bar \psi}} \gamma ^5{\hat \psi})$ or 
${\hat \psi} \gamma ^{\mu} 
({\hat {\bar \psi}} \gamma _{\mu} {\hat \psi})$ or 
${\hat \psi} \gamma ^{\mu} \gamma 
^5 ({\hat {\bar \psi}} \gamma _{\mu} \gamma ^5{\hat \psi} )$. 
The constant 
$l$ has units of length, and sets the scale for the strength of the 
interaction. Heisenberg emphasized that the 2-point Green's function,
$G_2 (x_2, x_1)$, in this theory differs strongly from the propagator in 
a linear theory in its behavior on the light
cone : in the nonlinear theory $G_2 (x_2 , x_1)$ oscillates strongly on
the light cone in contrast to the propagator of the linear theory
which has a $\delta$-like singularity. Heisenberg defined 
$\tau$ functions 
\begin{equation}
\label{h2}
\tau (x_1 x_2 ... | y_1 y_2 ...) = \langle 0 | 
T[{\hat \psi} (x_1) {\hat \psi} (x_2) ...
{\hat {\bar \psi }} (y_1) {\hat {\bar \psi }} (y_2) ...] | 
\Phi \rangle
\end{equation}
where $T$ is the time ordering operator; $| \Phi \rangle$ is a state for
the system described by Eq. (\ref{h1}). Eq. (\ref{h2}) 
establishes a one-to-one correspondence between the system state,
$| \Phi \rangle$, and the set of functions $\tau$. This state can be defined
using the infinite function set of Eq. (\ref{h2}). Using
equation (\ref{h1}) and (\ref{h2}) we obtain the following infinite
system of equations for the $\tau $'s
\begin{eqnarray}
\label{h3}
l^{-2} \gamma ^{\mu} _{(r)} {\partial \over \partial x^{\mu} _{(r)}}
&&\tau (x_1 ...x_n |y_1 ... y_n ) = \Im [ \tau (x_1 ... x_n x_r |
y_1 ... y_n y_r)] + \nonumber \\ 
&&\delta (x_r -y_1) \tau ( x_1 ... x_{r-1} x_{r+1} ... x_n |
y_2 ... y_{r-1} y_{r+1} ... y_n ) + \nonumber \\
&&\delta (x_r - y_2) \tau (x_1 ... x_{r-1} x_{r+1} ... x_n |
y_1 y_2 ... y_{r-1} y_{r+1} ... y_n ) + ...
\end{eqnarray}
Eq. (\ref{h3}) represents one of an infinite set of coupled equations 
which relate various orders (given by the index $n$) of the $\tau$ 
functions to one another. To make some head way toward solving 
this infinite set of equations Heisenberg employed the Tamm-Dankoff
method whereby he only considered $\tau$ functions up to a certain
order. This effectively turned the infinite set of coupled equations 
into a finite set of coupled equations.
\par 
Heisenberg used the procedure sketched above to study the
Dirac equation with a nonlinear coupling. Here we apply this
procedure to nonlinear, bosonic field theories such as QCD in
the low energy limit. In particular by
applying this method to the
infinite energy, solutions just discussed, and making certain
assumptions analogous to the Tamm-Dankoff cut-off,
the unphysical asymptotic behavior of these classical Yang-Mills 
solutions is ``smoothed'' out. It is also possible that
this quantization method could be applied to the
recently proposed \cite{dzh0} strongly interacting
phonon theory of high-$T_c$ superconductors.

\section{Quantization of singular classical solutions}

Before applying the above quantization method
we make the following simplifying assumptions :
\begin{enumerate}
\item
The degrees of freedom relevant for studying the 
singular solutions (both classically and quantum mechanically) 
are given entirely by the two ansatz functions $f ,v$ for 
Eqs. (\ref{18}) or,  $v,w$ for Eqs. (\ref{9}) or $f, \varphi$ 
for Eqs. (\ref{monp1}). No other
degrees of freedom arise through the quantization process.
\item
>From Eqs. (\ref{21}), (\ref{11}) and (\ref{monp4}) one function is 
a smoothly varying function for large
$x$, while another function is strongly oscillating. 
We take the smoothly varying function to be
an almost classical degree of freedom while 
the oscillating function is treated as a
fully quantum mechanical degree of freedom. 
Naively one might think that
in this way only the behavior of second function 
would change while first function stayed
the same. However since both functions are interrelated 
due to the nonlinear nature of the field equations 
we find that both functions are modified. 
\end{enumerate} 

\subsection{Pedagogical example : Weak anharmonic 
oscillator}

As a pedagogical example we repeat some details
of Heisenberg's quantization method as applied
to the weak anharmonic oscillator \cite{hs1}. This section
is also intended as a reference for when we apply
this same procedure to the various classical
Yang-Mills solutions. The Hamiltonian equations for
the operators of the weak anharmonic oscillator are
\begin{mathletters}
\label{ped1}
\begin{eqnarray}
{\hat {\dot q}} & = & \frac{{\hat p}}{m}, 
\label{ped1:1}\\
\frac{\hat {\dot p}}{m} & = & 
-\omega ^2_0 {\hat q} - \lambda {\hat q^3}
\label{ped1:2}
\end{eqnarray}
\end{mathletters}
here ${\hat q}$ and ${\hat p}$ are the operators 
of the canonical conjugate variables for the anharmonic 
oscillator. The $\tau$ functions are
\begin{equation} 
\tau (k|n-k) e^{i\omega t} = \left .\langle 0 | T q(t_1)q(t_2) 
\cdots q(t_k) 
p(t_{k+1}) p(t_{k+2})\cdots p(t_n)| \Phi\rangle 
\right |_{t_1 = t_2 = \cdots = t_n = t}
\label{ped2}
\end{equation}
The Hamiltonian operator equations yield the following
system: 
\begin{eqnarray}
i\omega \tau (k|l) = \frac{k}{m} \tau (k-1|l+1) - 
m\omega ^2_0 l \tau(k+1|l-1) - 
\nonumber \\
m\lambda l \tau (k+3|l-1) + 
m\lambda \hbar ^2 \frac{l(l-1)(l-2)}{4} \tau (k+1|l-3) 
\label{ped3} 
\end{eqnarray}
To a first approximation this reduces to : 
\begin{mathletters}
\label{ped4}
\begin{eqnarray}
i\omega \tau (1|0) & = & \frac{1}{m} \tau (0|1), 
\label{ped4:1}\\ 
\frac{i \omega}{m} \tau (0|1) & = & -
\omega ^2_0 \tau (1|0) - \lambda \tau (3|0)
\label{ped4:2}
\end{eqnarray}
\end{mathletters}
This is not a closed system, but by making the
following approximation:
\begin{equation} 
\tau (3|0) = \frac{3}{2}\frac{\hbar}{m\omega^2_0} \tau (1|0), 
\label{ped5}
\end{equation}
one can solve Eqs. (\ref{ped4}) to obtain: 
\begin{equation} 
\omega = \pm \omega _0\left ( 1 + \frac{3}{4} 
\frac {\hbar \lambda}{m\omega _0^3} \right ).
\label{ped6}
\end{equation}
Which coincides with the quantum mechanical 
solution as an expansion of order $\lambda$ : 
$\omega _{1,0} = (E_1 - E_0)/\hbar$ 

\subsection{The quantized SU(2) gauge ``string''}

To apply Heisenberg's quantization scheme to the Yang-Mills
system of Eq. (\ref{18})
we replace the ansatz functions by operators ${\hat f} (\rho) ,
{\hat v} (\rho )$ as in \cite{dzh2}: 
\begin{mathletters}
\begin{eqnarray}
\label{s9}
{\hat f}'' + {{\hat f}' \over x} &=& {\hat f} {\hat v}^2
\label{s9:1} \\
{\hat v}'' + {{\hat v}' \over x} &=& -{\hat v} {\hat f}^2
\label{s9:2}
\end{eqnarray}
\end{mathletters}
the primes denote derivatives with respect to 
the dimensionless radius $x$. Taking into
account assumption (2) we let ${\hat f} \rightarrow f$ become just
a classical function again, and replace ${\hat v}^2$ 
in Eq. ({\ref{s9:1}) by its expectation value
\begin{mathletters}
\begin{eqnarray}
\label{s10}
f'' + {f' \over x} &=& f \langle v^2 \rangle 
\label{s10:1} \\
{\hat v}'' + {{\hat v}' \over x} &=& - {\hat v} f^2 
\label{s10:2}
\end{eqnarray}
\end{mathletters}
Now if we took the expectation value of Eq. (\ref{s10:2}) and
ignored the coupling to $f$ on the right hand side we would
have an equation for determining $\langle v \rangle =
\langle \Phi |{\hat v} | \Phi \rangle$. However the two
nonlinear terms on the right hand side of Eqs. (\ref{s10:1} - \ref{s10:2}) 
show that a new object, $\langle v^2 \rangle$ enters the picture so that
Eqs. (\ref{s10:1} - \ref{s10:2}) are not closed. To obtain an
equation for $\langle v^2 \rangle$ we act on ${\hat v} ^2 (x)$
with the operator 
$\left( {d^2 \over dx^2} + {1 \over x} {d \over dx} \right)$ giving
\begin{equation}
\label{s11}
({\hat v}^2)'' + {1 \over x} ({\hat v}^2)' = -2 {\hat v}^2 f^2
+ 2 ({\hat v}')^2
\end{equation}
Taking the expectation value of this equation gives the desired
equation for $\langle v^2 \rangle$
\begin{equation}
\label{s12}
\langle v^2 \rangle '' + {1 \over x} \langle v^2 \rangle ' =
-2 \langle v^2 \rangle f^2 + 2 \langle (v') ^2 \rangle
\end{equation}
Again this equation is not closed due to the $\langle (v') ^2 \rangle$ 
term. We could again try to find an equation for $\langle
(v') ^2 \rangle$ by the same procedure we employed for $\langle
v^2 \rangle$. This equation would also not be closed. Continuing
in this way we would find an infinite set of equations. In order
to have some hope of handling this problem we need to make some
approximation to cut this process off.
For the anharmonic oscillator Heisenberg solved this
by assuming that $\tau (\underbrace {xx\cdots x}_{m} | 
\underbrace {pp\cdots p}_{n} ) = 
\underbrace {\langle x(t)x(t)\cdots x(t)}_{m} 
\underbrace {p(t)p(t)\cdots p(t)}_{n} \rangle 
\approx 0$ for large $m$ or $n$. 
We try two different approximations for the $\langle (v') ^2 \rangle$
term and show that both yield similar large $x$ behavior that 
fixes the infinite field energy problem of the classical solution.
First we assume that $\langle (v') ^2 \rangle \approx
\pm \langle v^2 \rangle '$. Since $\langle (v') ^2 \rangle$
is positive definite one picks the $\pm$ sign so that the
right hand side of this assumption is also positive definite.
Under this assumption the equations become
\begin{mathletters}
\begin{eqnarray}
\label{s16}
\langle v^2 \rangle '' +\left( {1 \over x} \mp 2 \right) \langle
v^2 \rangle ' &=& - 2 \langle v^2 \rangle f^2 
\label{s16:1} \\
f'' + {1 \over x} f' &=& f \langle v^2 \rangle
\label{s16:2}
\end{eqnarray}
\end{mathletters}
The approximate solution of Eqs. (\ref{s16:1} - \ref{s16:2}) 
is of the form
\begin{mathletters}
\begin{eqnarray}
\label{s14}
\langle v^2 \rangle & \approx & v_0 ^2 {exp (-\gamma x) \over \sqrt{x}} 
\label{s14:1} \\
f & \approx & f_{\infty} + f_0 {exp(-\gamma x) \over \sqrt{x}}
\label{s14:2}
\end{eqnarray}
\end{mathletters}
where 
\begin{equation}
\label{s17}
f_0 \gamma ^2 = f_{\infty} v_0 ^2 \; \; \; \; \; \; \;
\gamma ^2 \pm 2 \gamma = - 2 f_{\infty} ^2
\end{equation}
The second relationship can be written (using the 
first relationship) as
\begin{equation}
\label{s18}
\gamma = \mp f_{\infty} \left( f_{\infty} + {v_0 ^2 \over 2 f_0} \right)
\end{equation}
Although $\langle v'^2 \rangle \approx + \langle v^2 \rangle '$
leads to unphysical exponentially growing solutions, the
assumption $\langle v'^2 \rangle \approx - \langle v^2 \rangle '$ leads
to exponentially decaying solutions. Thus under this latter assumption
we find that the quantum mechanical treatment
of this nonlinear system modifies the bad features of the classical
solution. The asymptotic behavior of $v$ goes from being strongly
oscillating (see Eq. (\ref{21:2})) to decaying exponentially, while the
asymptotic behavior of $f$ goes from being linearly increasing 
(see Eq. (\ref{21:1})) to also decaying exponentially.
In some intermediate region the two solutions should match
up with one another. If the asymptotic
form for these ansatz functions are used in the energy density, ${\cal E}$,
of Eq. (\ref{21:3}) we find that the field energy is now finite. To
calculate ${\cal E}$ we would replace the classical terms $v'^2 , f^2 v^2$ 
by the appropriate quantum operator and take the expectation value.
The $\langle v'^2 \rangle$ term would be handled according to the assumption
we used for closing the equations.

\subsection{Quantized ``bunker'' solution}
\label{bunker} 

Although the classical behavior of ``bunker'' solution
is interesting due to its similarity with certain phenomenological
confining potentials, the infinite field energy discussed at the end of
section \ref{su3} strongly argues against the physical importance
of this solution. One possible escape from this conclusion is if
quantum effects weakened or removed this bad long distance behavior. 
\par 
Again in Eq. (\ref {9})
we replace the ansatz functions by operators ${\hat w} (x) ,
{\hat v} (x)$ \cite{dzh3}: 
\begin{mathletters}
\begin{eqnarray}
\label{b9}
x^2 {\hat v}'' &=& {\hat v}^3 - {\hat v} - {\hat v} {\hat w} ^2
\label{b9:1} \\
x^2 {\hat w}'' &=& 6{\hat w} {\hat v}^2
\label{b9:2}
\end{eqnarray}
\end{mathletters}
here the prime denotes derivatives with respect to 
the dimensionless radius $x$. Taking into
account assumption (2) we let ${\hat w} \rightarrow w$ become just
a classical function again, and replace ${\hat v}^2$ 
in Eq. ({\ref{b9:2})
by its expectation value to arrive at
\begin{mathletters}
\begin{eqnarray}
\label{b10}
x^2 {\hat v}'' &=& {\hat v}^3 - {\hat v} - {\hat v} w ^2 
\label{b10:1} \\
x^2 w'' &=& 6 w \langle v^2 \rangle
\label{b10:2}
\end{eqnarray}
\end{mathletters}
If we took the expectation value of Eq. 
(\ref{b10:1}) we would almost have a closed system of differential
equations relating $w$ and $\langle {\hat v} \rangle$. 
The $\langle {\hat v}^2 \rangle$
term from Eq. (\ref{b10:2}) and the $\langle {\hat v}^3 \rangle$ term
from Eq. (\ref{b10:1}) prevent the equations from being closed.
Applying the operation $x^2 \partial ^2 / \partial x ^2$ to the operator
${\hat v} ^2$ and using Eq. (\ref{b10:1}) yields
\begin{equation}
\label{b11}
x^2 ({\hat v ^2}) '' = 2 {\hat v} ^2 ({\hat v} ^2 -1 -w^2)
+ 2 x^2 ({\hat v}') ^2
\end{equation}
If we took the expectation of the above equation with respect
to fluctuations in the ansatz function operator ${\hat v^2}$,
and combined this with Eq. (\ref{b10:2}) we would almost have
a closed system for determining $w$ and ${\hat v^2}$ except for
the $\langle ({\hat v}')^2 \rangle$ term
on the right hand side of Eq. (\ref{b11}). Continuing
in this way one could obtain an infinite set of equations for
various powers of the ansatz function operator ({\it i.e.} 
${\hat v ^n}$). As in the previous
section we make some assumption that effectively cuts
off the system of equations at some finite order. 
We assume that the mean square of the $\phi$ component 
of the magnetic field ${\hat H_\phi} = {\hat v}'$ is small 
\footnote{in other words the fluctuations of 
the ${\hat H_\phi}$ component of the magnetic field are weak.} 
\begin{equation}
\label{b11a}
\left\langle (\Delta\hat H_\phi) ^2 \right\rangle = 
\left\langle \left ({\hat H_\phi} - 
\left\langle {\hat H_\phi}\right\rangle \right )^2 \right\rangle = 
\left\langle {\hat H_\phi}^2 \right\rangle - 
\left\langle {\hat H_\phi}\right\rangle ^2 \approx 0
\end{equation}
that implies 
\begin{equation}
\label{b12}
\left\langle {\hat H_\phi}^2 \right\rangle \approx 
\left\langle {\hat H_\phi}\right\rangle ^2 
\quad {\mbox or } \quad 
\left\langle ({\hat v'}) ^2 \right\rangle \approx 
(\left\langle {\hat v'} \right\rangle ) ^2
\end{equation}
Then by taking the expectation of Eq. (\ref{b11}) 
we arrive at a closed system of equations from
Eqs. (\ref{b10:1}) (\ref{b10:2})  (\ref{b11})
\begin{mathletters}
\begin{eqnarray}
\label{b13}
x^2 \langle {\hat v}^2 \rangle '' &=& 2 \langle {\hat v}^2 \rangle ^2
-2 \langle {\hat v}^2 \rangle w^2 - 2 \langle {\hat v}^2 \rangle
+2x^2 (\langle {\hat v} \rangle ') ^2
\label{b13:1} \\
x^2 w'' &=& 6 w \langle {\hat v}^2 \rangle
\label{b13:2} \\
x^2 \langle {\hat v} \rangle '' &=& \langle {\hat v} \rangle ^3
- \langle {\hat v} \rangle - w^2 \langle {\hat v} \rangle
\label{b13:3}
\end{eqnarray}
\end{mathletters}
here we have further assumed that
$\langle v^4\rangle  \approx \langle v^2\rangle ^2$
and $\langle v^3 \rangle  \approx \langle v \rangle ^3$.
Note that the present situation is slightly more
complex than the quantization of the string solution
of the previous section since now we have three coupled
equations rather than two.
It is straightforward to show that in the limit $x \rightarrow
\infty$ the closed system given by Eqs. (\ref{b13:1}) - (\ref{b13:3})
is solved by
\begin{mathletters}
\begin{eqnarray}
\label{b14}
\langle {\hat v}^2 \rangle &\approx& 1  + {a^2 \over 2 x ^2}
\label{b14:1} \\
w &\approx& {b \over x^2}
\label{b14:2} \\
\langle {\hat v} \rangle &\approx& \pm 1 + {a \over x}
\label{b14:3}
\end{eqnarray}
\end{mathletters}
Eqs. (\ref{b14:1}) - (\ref{b14:3}) 
provide information about the behavior of the ``classical'' ansatz
function $w$, and the ``quantum'' ansatz function, $v$,
via $\langle {\hat v} \rangle$ and $\langle {\hat v}^2 \rangle$. 
The main point is that after applying the 
Heisenberg-like quantization procedure to the classical 
singular solution, the infinite increase of
the ansatz function $w$, has changed to an acceptable
asymptotic behavior ({\it i.e.} one that leads to
a finite field energy). By replacing $v^2 , w, (v')^2$
in Eq. (\ref{14}) with $\langle {\hat v}^2
\rangle , w, (\langle {\hat v} \rangle ') ^2$ -- from Eqs. 
(\ref{b14:1}) - (\ref{b14:3}) -- we
find that the field energy density of the quantized ``bunker'' 
solution takes the form
\begin{equation}
\label{b16}
{\cal E} \propto {1 \over r^6}
\end{equation}
in the limit in which quantum fluctuations become important
({\it i.e.} for non-Abelian theories which exhibit
asymptotic freedom this means in the low energy or
$r \rightarrow \infty$  range) the energy density goes from
the form given in Eq. (\ref{14}) to that given in Eq. (\ref{b16}).
This gives a finite field energy. In the high
energy or short distance regime we assume that the 
fields approach the classical configuration of Ref. \cite{dzh1} 
due to asymptotic freedom. This classical configuration
is well behaved at $r=0$, but has an infinite
field energy due to its divergence as $r \rightarrow \infty$.

\subsection{The quantized $f, \varphi \neq 0$ case}
\label{monopole}

Finally, we employ the above quantization technique to the  
singular solutions of Eqs. (\ref{monp1}) for the $v=w=0$ case. 
Again we replace the ansatz functions by operators ${\hat f} (x) ,
{\hat \varphi} (x)$ 
\begin{mathletters}
\begin{eqnarray}
\label{mon1}
x^2 {\hat f}'' &=& {\hat f}^3 - {\hat f} - {\hat f} 
{\hat \varphi} ^2
\label{mon1:1} \\
x^2 {\hat \varphi}'' &=& 2{\hat \varphi} {\hat f}^2
\label{mon1:2}
\end{eqnarray}
\end{mathletters}
using assumptions similar to the previous sections we have the 
following equations for the quantum function $f$, and the 
classical function $\varphi$: 
\begin{mathletters}
\begin{eqnarray}
\label{mon2}
x^2 {\hat f}'' &=& {\hat f}^3 - {\hat f} - {\hat f} 
\langle \varphi ^2 \rangle 
\label{mon2:1} \\
x^2 \varphi'' &=& 2 \varphi \langle f^2 \rangle
\label{mon2:2}
\end{eqnarray}
\end{mathletters}
Using the same assumptions (including the assumption about the weak 
fluctuation) and procedure as in the previous
sub-section we can turn Eqs. (\ref{mon2}) into a closed system
of equations for the expectation values of various powers
of the ansatz function operators
\begin{mathletters}
\begin{eqnarray}
\label{mon2a}
x^2 \langle {\hat f}^2 \rangle '' &=& 2 \langle {\hat f}^2 \rangle ^2
-2 \langle {\hat f}^2 \rangle \varphi ^2 - 2 \langle {\hat f}^2 \rangle
+2x^2 (\langle {\hat f} \rangle ') ^2
\label{mon2a:1} \\
x^2 \varphi'' &=& 2 \varphi \langle {\hat f}^2 \rangle
\label{mon2a:2} \\
x^2 \langle {\hat f} \rangle '' &=& \langle {\hat f} \rangle ^3
- \langle {\hat f} \rangle - \varphi ^2  \langle {\hat f} \rangle
\label{mon2a:3}
\end{eqnarray}
\end{mathletters}
This system has the following asymptotic solution
\begin{mathletters}
\label{mon3}
\begin{eqnarray}
\langle f^2\rangle  & \approx & 1 - {a \over x}, 
\label{mon3:1}\\
\langle f\rangle  & \approx & \pm 1 - {a \over x} , 
\label{mon3:2}\\
\varphi & \approx & {b \over x}
\label{mon3:3}
\end{eqnarray}
\end{mathletters}
The field energy density for this solution takes the form 
\begin{equation}
\label{mon4}
{\cal E} \propto {1 \over r^6}
\end{equation}

\section{Operator quantization for strongly interacting fields}

In the previous sections we saw that strongly nonlinear 
fields ({\em e.g.} non-Abelian gauge fields) 
can have nonlocal objects (the classical SU(2) flux
tube solution and the SU(2),SU(3)``bunker'' solutions) which
have good long distance behavior after an application
of the Heisenberg quantization procedure. 
This result is markedly different 
from the situation for linear quantum field 
theories with no or small self-interaction. In these cases 
the quantum fields can not form static field
configurations such as monopoles, flux tubes, or
``bunker'' solutions discussed above. In general
one needs strongly interacting, nonlinear theories
in order for such objects to exist.
\par
All of this may be an indication that
\textit{\textbf {the n-point Green's functions in
non-Abelian gauge theories can 
have some pieces which are not simply the sum of Feynman 
diagrams}}. Physically, this would imply that some physical 
processes in non-Abelian gauge theories can not be 
explained solely in terms of the perturbative fluctuations of
field quanta. Heisenberg may have had this situation in mind when
he indicated \cite{hs1} that the divergence behavior in a theory 
with strong interactions will differ from that found
in a perturbatively, renormalizable theory like quantum
electrodynamics. In this section we want to first present some
formal and then some heuristic arguments which support this idea
of non-perturbative, non-Feynman contributions to the
Green's functions.

\subsection{Quantum field theory with nonassociative field operators} 

In this sub-section we consider field operators which are
nonassociative. There are two reasons for doing this.
First, in the Heisenberg quantization scheme one
problem (which was dealt with by an approximate Tamm-Dankoff
cutoff) was the appearance of an infinite system of differential
equations with interrelated Green's functions.
Nonassociative field operators can lead to a quantum theory
where the Green's functions are functionally independent
of one another, which would help this problem. Second,
a nonassociative quantum theory can lead to amplitudes
and processes which cannot be represented simply
in a Feynman diagrammatic way by the exchange
of virtual quanta. This supports the
general picture given in the introduction to this
section. In the following we will use the notations
and conventions of Ref. \cite{dzh4}.
\par
In the standard, noncommutative quantum field theories 
it is impossible to define an n-point Green's function that is not some
polylinear combination of propagators and interaction vertices.
This general feature of the Green's functions is
altered if the field operators are nonassociative. If one
considers the time-ordering of a product of $n$ field operators
then the time-ordering will result in a shifting of the positions
of the individual field operators. In a noncommutative theory
the shifting in the positions of the field operators usually
leads to the appearance of a commutator. The time-ordering can
also lead to a change in the ordering of the brackets which
group field operators together. In a nonassociative theory
such a change in the ordering of the brackets leads to the
appearance of an {\it associator}, just as in a
noncommutative algebra the displacement of an operator to the left
or right gives rise to a commutator. A good feature of
nonassociative quantum field theories is that one can
obtain n-point Green's functions that are independent of each other.
\par
We denote the product of $n$ field operators
in which the brackets are arranged in accordance with some rule
$P$ by $M_n(P)$. For example,
\begin{equation}
M_3(P) =\Bigl ((\hat \varphi _x \hat \varphi _y )\hat \varphi _z\Bigl ),
\label{a1}
\end{equation}
where $\hat\varphi _x = \hat\varphi (x)$
is the field operator at the point $x$;
in this case, the rule $P$ is that all opening brackets are at the
extreme left hand position.
\par
We call this ordering of the brackets for any number of operators,
( {\it i.e.} with all opening brackets in the extreme left
hand position) the normal position ordering of the brackets,
and denote it by colons
\begin{equation}
: M_n(P) := \Biggl (\biggl (\cdots \Bigl (\hat \varphi _1(
\hat\varphi _2 \cdots \biggl )\hat\varphi _n\Biggl ),
\label{a2}
\end{equation}
On the left there are $n$ opening brackets in succession.
\par
Next, we consider a nonassociative quantum field theory with
the following axioms.
\par
1. Two monomials $M_n(P_1)$ and $M_n(P_2)$ that differ only in the
placement rule of the brackets differ from each other by a
numerical function $A(x_1,x_2,\ldots ,x_n; P_1, P_2)$,
which we call the n-point associator:
\begin{equation}
M_n(P_1) - M_n(P_2) = A(x_1,x_2,\ldots ,x_n;P_1,P_2).
\label{a3}
\end{equation}
\par
2. The action of the product of two monomials on the quantum state
vector $|v\rangle $ is defined as follows:
\begin{mathletters}
\begin{eqnarray}
\label{a4}
\Bigl ( M_n(P_1) M_k(P_2) \Bigl ) |v\rangle  & = &
M_n(P_1) \Bigl (M_k(P_2)|v\rangle  \Bigl ),
\label{a4:1}\\
\langle v| \Bigl ( M_n(P_1) M_k(P_2) \Bigl ) & = &
\Bigl ( \langle v| M_n(P_1) \Bigl ) M_k(P_2).
\label{a4:2}
\end{eqnarray}
\end{mathletters} 
\par
3. We have the usual commutation rules
\begin{equation}
\left [ \hat \varphi _x , \hat \varphi _y \right ] = G(x,y),
\label{a5}
\end{equation}
where $G(x, y)$ is the 2-point Green's function (propagator) 
which differs from the propagator of a linear theory. 
\par
As an example of this formalism we examine the following
simplified model, in which the action of any field operator on the
vacuum state annihilates it
\begin{equation}
\hat\varphi _x |vac\rangle  = \langle vac|\hat\varphi _x = 0.
\label{a6}
\end{equation}
Now the 3-point associator is given by
\begin{equation}
(\hat\varphi _x\hat\varphi _y)\hat\varphi _z =
\hat\varphi _x(\hat\varphi _y\hat\varphi _z) +
A(x,y,z).
\label{a7}
\end{equation}
Using the rules introduced above for the action of the field operators
on the vacuum state, we find that $A(x,y,z)=0$.
There are two 4-point associators given by 
\begin{mathletters}
\begin{eqnarray}
\label{a8} 
\Bigl ((\hat\varphi _x\hat\varphi _y)\hat\varphi _z\Bigl )\hat\varphi _u
& = & (\hat\varphi _x\hat\varphi _y)(\hat\varphi _z\hat\varphi _u) +
A_1(x,y|z,u),
\label{a8:1}\\
\hat\varphi _x\Bigl (\hat\varphi _y(\hat\varphi _z\hat\varphi _u)\Bigl )
& = & (\hat\varphi _x\hat\varphi _y)(\hat\varphi _z\hat\varphi _u) +
A_2(x,y|z,u),
\label{a8:2}
\end{eqnarray}
\end{mathletters}
Acting on each of Eqs.(\ref{a8}) from the right and left with
the vacuum state and using the condition (\ref{a6}), the 4-point
associators $A_{1,2}$ can be expressed in terms of the vacuum
expectation value of the monomial
$(\hat\varphi _x\hat\varphi _y)(\hat\varphi _z\hat\varphi _u)$:
\begin{equation}
A_{1,2}(x,y|z,u) = A(x,y|z,u) = -\langle vac|
(\hat\varphi _x\hat\varphi _y)(\hat\varphi _z\hat\varphi _u)|vac\rangle .
\label{a9}
\end{equation}
$x^{\mu}\sim y^{\mu},z^{\mu}\sim u^{\mu}$
($x^{\mu}, y^{\mu}, z^{\mu}$ and $u^{\mu}$
are 4-vectors). Using the commutation properties
of the operators, the property of the vacuum state, and taking the
points to be pairwise spacelike
($x^{\mu}\sim y^{\mu},z^{\mu}\sim u^{\mu}$) we can establish
the following symmetry properties of the 4-point associator:
\begin{equation}
A(x,y|z,u) = A(y,x|z,u) = A(x,y|u,z).
\label{a10}
\end{equation}
The Green's function can now be defined as
\begin{equation}
T(M_n(P)) = M_n(P) + G(x_1,x_2,\ldots ,x_n;P),
\label{a11}
\end{equation}
where $T$ denotes the ordinary time ordering of the operators
in the monomial $M_n(P)$, and also normal ordering of
the brackets in the monomial $M_n(P)$. Thus each 
n-point Green's function can be expressed in terms of a polylinear
combination of $n$-point associators $(m\le n)$ and propagators.

\subsection{Heuristic argument for non-Feynman contributions
to the Green's functions} 

In carrying out our approximate application of 
Heisenberg's quantization method to certain classical, singular 
solutions, we examined only two degrees of freedom 
($f$ and $v$,  $v$ and $w$ or $f$ and $\varphi$). 
In each of these pairs one function 
was taken as a quantum degree of freedom and the other remained a classical 
degree of freedom. To investigate this method further one
could drop this assumption by treating both functions as
quantum degrees of freedom. In this section we
look at a broader issue of how to include the other 
frozen degrees of freedom. In other words the quantization
process might lead to more degrees of freedom beyond the
original classical degrees ({\it e.g.} $f$ and $v$).
\par 
Our main assumption in addressing this question is that
for any n-point Green's function, for strongly interacting,
nonlinear fields, there is a non-perturbative piece
which can be treated by the Heisenberg quantization method 
\footnote{this piece may be connected with the nonassociative properties 
of fields operators}. 
In the classical region these degrees of freedom region can be
associated with classical (possibly singular) field configurations.
The others degrees of freedom are then handled 
using ordinary perturbative methods ({\it i.e.} using Feynman
diagram techniques). 
\par 
Physically, this means that in a theory with 
strongly interacting fields one has: 
\begin{enumerate} 
\item
nonlocal static objects, formed by those quantum degrees 
of freedom which are connected with (possibly singular)
solutions of the classical field equations. 
\item 
the quanta which fluctuate around 
these nonlocal objects.
\end{enumerate} 
\par 
In terms of QCD this implies that there are 
degrees of freedom which play a dominate role
in the Feynman path integral
to a first approximation. These degrees of freedom 
are connected with classical (possibly singular) 
solutions of the Yang-Mills field equations. This is 
somewhat similar to the saddle-point method of evaluating the 
Feynman path integral.  However, in the 
saddle-point calculation one approximates
the path integral via the value of the action
evaluated at a particular solution
\begin{equation} 
\int D\varphi e^{S[\varphi]} \approx 
e^{S[\varphi _0]}
\label{d1}
\end{equation} 
here $\varphi$ is a field degree of freedom (in our case it is
the non-Abelian potential $A^a_\mu$), and $\varphi _0$ is 
a classical, nonsingular solution of the field equations. 
For singular, classical solutions this approximation is not valid.
Instead one can as a first approximation take 
\begin{equation} 
\int D\varphi e^{S[\varphi]} \approx 
\int D\varphi _{cl} e^{S[\varphi_{cl}]}
\label{d2} 
\end{equation} 
here $\varphi _{cl}$ represents the ansatz functions
used in the classical field equation. For
the ansatz functions of Eq. (\ref{3})
($f,v,w,\varphi$) the integration would become
$D\varphi _{cl} \rightarrow Df Dv Dw D\varphi$.
This leads to the following
expression for the path integral
\begin{equation} 
\int D A^a_\mu e^{S[A^a_\mu]} =
\int D (A^a_\mu)_{dev} \left [
\int D (A^a_\mu)_{cl} e^{iS[{(A^a_\mu)}_{cl} + 
{(A^a_\mu)}_{dev} ]} \right ]
\label{d3} 
\end{equation} 
here ${(A^a_\mu)}_{cl}$ are the classical solutions
in terms of the ansatz functions,
and ${(A^a_\mu)}_{dev}$ are the perturbative, quantum
deviations from the classical solution. The first integral 
is calculated by some nonperturbative 
method (lattice gauge calculation, Heisenberg method {\it etc.}).
The second path integral can be handled
by ordinary perturbative methods (Feynman diagrams). 
\par
At this point one can ask how renormalization
fits in with the above development where
the Green's functions contain a non-perturbative
piece. In the standard perturbative treatment
of a quantum field theory one must introduce
a renormalization prescription. The physical 
basis for this can to some extent be traced
to the point-like nature of the quanta and
interaction vertices. Mathematically the
divergences come from momentum integrals
over closed loops which have propagators
(which are distributions in linear field theories)
raised to some power. In the developments
of the previous sections, where the n-point Green's
functions have a non-perturbative part due to the
non-linear nature of the quantized fields, it is no
longer necessarily true that the Green's functions
need to be distributions. A generalization of the
standard Feynman diagram technique would represent
any complete n-point Green's functions
as the sum of two pieces : one the standard,
perturbatively calculated field fluctuations, and the
other would be the non-Feynman part (possibly related to
nonsingular or weakly singular classical solutions).
A similar situation arises in superstring theory
where the building blocks for the scattering 
diagrams are nonlocal strings. In both the present case and in the
case of string theory the divergences of the quantum theory
are eliminated or decreased  
\footnote {A related question would be to investigate
whether there is a classical, nonlinear field theory which 
after quantization resulted in n-point Green's functions 
with non-perturbative pieces that had a structure similar to
$p$-branes. If so then one might consider string theory 
as a quantum field theory arising from some nonlinear 
(possible nonlocal) field theory !}.  

\section{Discussion and conclusions}

The basic idea proposed here is that 
\textit{\textbf{quantum field theories 
with strongly non-linear fields can have the non-Feynman pieces.}} 
Also the classical, possibly singular solutions can play
an important role in the corresponding quantum field theory.
After quantization these classical solutions can lead to nonlocal 
objects with good long distance behavior. These quantized nonlocal 
object are connected with the nonperturbative part of the
Green's functions, and they are not the result of the perturbative
quantum fluctuations of the field.
\par
This approach to the quantization of non-Abelian
field theories is very similar to the attempts
to quantize gravity on the basis of the Wheeler-de Wit equation.
In the present case there is an important distinction :
the operator of canonical momentum is not simply
related to the derivative of a canonical coordinate
({\it i.e.} ${\hat {\cal P}} \neq {{\delta}\over{\delta \dot q}}$ 
where ${\hat q}$ is a field operator). This follows from 
the fact that the propagator has an unusual behavior 
on the light cone. Reversing the analogy between the
non-Abelian field equations and the Wheeler-de Wit equation
one could postulate that the Wheeler-de Wit equation
may be more complicated than is normally written, 
since the momentum operator in gravity is also not simply
the ordinary derivative with respect to some corresponding
field variable.
\par 
In sections \ref{bunker} and \ref{monopole} we found 
two spherically symmetric nonlocal objects with 
good asymptotic behavior after quantization.
It may be possible to forge a connection between
these solutions and the QCD picture for the
``dressing''of valence quarks by gluons and 
quark-antiquark pairs. The authors of Ref. \cite{lavelle} 
write: ``One views dressing as surrounding 
the charged particle with a cloud of gauge fields. 
We $\ldots$ see that, $\ldots$ this cloud spreads out 
over the whole space, resulting in a highly nonlocal 
structure. This is not unexpected in QED, but 
in QCD we will see that there is a nonperturbative 
obstruction to the construction of this dressing.'' 
The Heisenberg quantization method that we have outlined
might provide a nonperturbative description of
the above ``dressing'' procedure. If this is correct then 
our results give structures which are ``dressed'' 
with gluons but not with quark-antiquark 
pairs since we did not consider quark fermion 
fields in this work. 

\subsection{Connection with the Maximal Abelian gauge model}
\label{maxag}

The quantization method of non-Abelian gauge fields
given in this work shares some
common features with the Abelian Projection 
(AP) model of gluodynamics proposed by 't Hooft 
in Ref. \cite{hooft}. The basic idea of the AP model is that
in a non-Abelian gauge theory such as QCD
there are special degrees of freedom which
play an essential role in confinement. These 
degrees of freedom are connected with monopoles which 
appear in the maximal Abelian subgroup, H, of the full
non-Abelian group G. In the AP model the initial 
SU(3) field equations are rewritten as an Abelian gauge 
theory with magnetic monopoles and some matter fields. 
Roughly speaking, the AP method implies that the 
off-diagonal gauge fields are excluded from consideration. 
More precisely one integrates over all the
off-diagonal (G/H) gauge fields, leaving only the
Abelian degrees of freedom \cite{kei}.
\par
SU(3) gauge theory has monopole solutions, which at large
distances look like U(1) Abelian monopoles ({\it i.e.}
the solution exhibits $1/r^2$ behavior asymptotically).
Thus after quantization these Abelian monopole solutions
should play a more important role in comparison
with the quantized solutions examined in sections
\ref{bunker} and \ref{monopole} 
since these quantized solutions have
a faster fall off. This is in agreement with the AP model.

\subsection{Connection with High-$T_c$ superconductivity}

The quantum theory of solids can be taken 
as a quantum field theory on some 
lattice. Thus one could ask if the above formalism
has an application to condensed matter systems.
For example can the quantized, classical solutions
discussed above play a role in condensed matter ? 
In Ref. \cite{obuch} it was shown that a flux tube solution 
can appear in superconductors in the mixed 
state. In \cite{dzh0} a model of Cooper pairing was
proposed in which the phonons exchanged between
Cooper electrons are confined in a ``flux'' tube
similar to the chromoelectric flux tubes which are
thought to occur between quarks and anti-quarks in
QCD. 
\par 
One possible application of such a model \cite{dzh5} 
might be as a mechanism to explain the Cooper pairing
in high-$T_c$ superconductors. One could postulate
that the Lagrangian for high-$T_c$ materials had a
strong, nonlinear potential term. This could result
in a classical, cylindrical solution (either singular
or nonsingular), which after quantization via
the Heisenberg-like method could give rise to a
phonon flux tube that stretched between two Cooper 
electrons, binding them to much higher
temperatures than in ordinary BCS superconductors.

\section{Acknowledgment} VD is grateful for financial support
from the Georg Forster Research Fellowship from the Alexander
von Humboldt Foundation and H.-J. Schmidt for an invitation to
Potsdam University.

\end{document}